\documentclass[aps,pra,showpacs,notitlepage]{revtex4-1}

\usepackage{graphicx}

\begin{document}

\title{Contextuality of quantum fluctuations characterized\\ by conditional weak values of entangled states}


\author{Holger F. Hofmann}
\email{hofmann@hiroshima-u.ac.jp}
\affiliation{
Graduate School of Advanced Science and Engineering, Hiroshima University,
Kagamiyama 1-3-1, Higashi Hiroshima 739-8530, Japan
}

\begin{abstract}
The quantum fluctuations of a physical property can be observed in the measurement statistics of any measurement that is at least partially sensitive to that physical property. Quantum theory indicates that the effective distribution of values taken by the physical property depends on the specific measurement context based on which these values are determined and weak values have been identified as the contextual values describing this dependence of quantum fluctuations on the measurement context. Here, the relation between classical statistics and quantum contextuality is explored by considering systems entangled with a quantum reference. The quantum fluctuations of the system can then be steered by precise projective measurements of the reference, resulting in different contextual values of the quantum fluctuations depending on the effective state preparation context determined by the measurement of the reference. The results show that mixed state statistics are consistent with a wide range of potential contexts, indicating that the precise definition of a context requires maximal quantum coherence in both state preparation and measurement. 
\end{abstract}

\maketitle

\section{Introduction}

Quantum states express the statistics of a physical property $\hat{A}$ as a superposition of eigentates of that physical property. As indicated by the Kochen-Spekker theorem \cite{KS}, these quantum fluctuations are contextual in the sense that their distribution of values depends on the precise measurement context that determines the individual values of $\hat{A}$. Recent research indicates that this context dependence of physical properties can be described by the weak values of the physical property that are defined by combining the initial state with the specific outcome obtained in the measurement \cite{Tol07,Dre10,Pus14}. There is a strong case to be made that weak values provide an accurate description of contextual quantum fluctuations because the magnitude of the fluctuations of weak values observed in pure states is exactly equal to the initial uncertainty of the observable for all possible measurement contexts \cite{Hos10,Hof11,Hal16,Hof20a}. If this understanding of contextual values is correct, a complete definition of contextual realities requires a pure state input representing the initial conditions and a pure state output representing the individual measurement result. Quantum states and quantum measurements each represent incomplete definitions of contexts, and this incompleteness is absolutely necessary to avoid logical contradictions within the formalism. At the fundamental level, quantum mechanics can then be understood as a description of the experimental possibilities of controlling and observing the changeable context of physical systems at the ultimate limit of precision. 

The similarity between classical fluctuations and the fluctuations of weak values determined in a quantum measurement has been discussed in great detail in \cite{Hal16}. One of the main insights from this analysis is that weak values provide a rather natural interpretation of operator statistics in terms of contextual estimates of the corresponding physical properties. In the following, I will develop this approach a bit further and take a closer look at the transition between the classical aspects of quantum fluctuations and their highly contextual limit. For this purpose, I consider the case of pure state entanglement between the system and a reference. When the reference is traced out, the statistics of the system is described by a mixed state and the contextuality of weak values can be reduced to the point where classical statistical arguments about the distribution of eigenvalues of $\hat{A}$ are sufficient to explain the weak values as conventional Bayesian estimates. However, quantum measurements of the reference provide additional information regarding the state preparation part of the context of $\hat{A}$, resulting in an unraveling of the quantum fluctuations that yields different weak values for different measurements of the reference. Importantly, the absence of signal transmission between the reference measurements and any information obtained from the system itself means that the statistics obtained by measurements of the system must be consistent with all possible state preparation contexts realized in this manner. Effectively, quantum entanglement permits the steering of the state preparation context, modifying the values of the optimal estimates of the property $\hat{A}$ by providing additional information about the initial conditions of the system. 

The analysis presented in the following highlights the relation between the partial definitions of context represented by state preparation and measurement. All state preparation procedures are consistent with all possible measurements, but the precise contexts defined by both initial conditions and measurement are not consistent with the precise context defined by any other combination. In the present scenario, weak value steering demonstrates that the same measurement can belong to very different contexts, depending on the precise quantum coherence of the input state. Entanglement is a non-local form of coherence, which means that the context of state preparation in the system will be given by the context decided by the measurement of the reference. Context can therefore be seen as an informational concept that is closely related to statistical correlations and their representation by fluctuations in the observed values of the various physical properties. Contextuality naturally emerges when the precision of control approaches the pure state limit. Any additional definition of context by measurements requires a disturbance of the previously established context, so that the definition of contexts by the initial conditions never achieves a completeness and precision that would be inconsistent with future measurements. The uncertainty principle is thus shown to be a necessary consequence of the separation of context into state preparation and measurement. 

The rest of the paper is organized as follows. In section \ref{Sec:context} the algebra of contextual statistics is reviewed and the difference between pure state and mixed state statistics is discussed. In section \ref{Sec:steering} it is shown how entanglement can be used to control the state preparation part of the contextuality of quantum fluctuations. In section \ref{Sec:relations} the relations between different contexts are discussed and the emergence of context independence in noisy statistics is explained. Conclusions are presented in section \ref{Sec:concl}.

\section{Quantum fluctuations in various contexts}
\label{Sec:context}

In general, a pure state $\mid \psi \rangle$ predicts the outcomes of a precise measurement of the physical property $\hat{A}$ with an uncertainty of $\Delta A$ given by the expectation values of the operator $\hat{A}$,
\begin{equation}
\Delta A^2 = \langle \psi \mid \hat{A}^2 \mid \psi \rangle - \langle \psi \mid \hat{A} \mid \psi \rangle^2.
\end{equation}
The most natural explanation of this formula is obtained by using the eigenstates $\mid a \rangle$ of the property $\hat{A}$ to relate the statistics of $\hat{A}$ to the probabilities of measurement outcomes $|\langle a \mid \psi \rangle|^2$. However, this is not the only possible decomposition of the operator algebra. In fact, it is difficult to justify the eigenvalue decomposition when the eigenvalues of $\hat{A}$ are not actually observed. The assignment of eigenvalues to the physical property $\hat{A}$ corresponds to a hidden variable theory, and various theorems of quantum physics show that such a measurement-independent assignment of hidden variables to local physical properties $\hat{A}$ is not consistent with the Hilbert space formalism. The most general proof that eigenvalues cannot be assigned independent of measurement context is given by the Kochen-Spekker theorem \cite{KS} which proves that the simultaneous assignements of eigenvalues to specific selections of non-commuting observables are necessarily inconsistent. The unavoidable conclusion must be that the assignement of values to fluctuating physical properties of a state $\mid \psi \rangle$ depends on the measurement context in which the fluctuations are observed. 

The quantum fluctuations of a physical property $\hat{A}$ are not just observed in direct measurements of that property, but also in measurements of any other orthogonal basis $\{ \mid m \rangle \}$. The specific outcome $\mid m \rangle$ is then decided by a function of the initial conditions expressed by the state $\mid \psi \rangle$ and the physical property given by the operator $\hat{A}$ \cite{Hof14,Nii18,Hof20b}. It is therefore possible to assign values of $A(m)$ to each measurement outcome by combining the information contained in the initial state $\mid \psi \rangle$ with the information gained from the measurement outcome $\mid m \rangle$. If this information is complete, the uncertainty $\Delta A$ can be expressed in terms of the contextual values $A(m)$ associated with each measurement outcome $\mid m \rangle$,
\begin{equation}
\label{eq:cqf}
\Delta A^2 = \sum_m |\langle m \mid \psi \rangle|^2 |A(m)|^2 - \left|\sum_m |\langle m \mid \psi \rangle|^2 A(m)\right|^2.
\end{equation}
The assignment of contextual values $A(m)$ to measurements that do not identify eigenvalues of the observable $\hat{A}$ should be governed by well-defined laws of physics expressed in terms of the operator algebra of quantum mechanics. It should therefore not be surprising that these contextual values have already been identified. As shown in previous research, the contextual values $A(m)$ of a physical property $\hat{A}$ are given by the complex weak values $A(m)$ defined by the initial state $\mid \psi \rangle$ and the measurement outcome $\mid m \rangle$ \cite{Tol07,Dre10},
\begin{equation}
\label{eq:WV}
A(m) = \frac{\langle m \mid \hat{A} \mid\psi \rangle}{\langle m \mid \psi \rangle}.
\end{equation}
For the present discussion, it is not relevant whether these weak values can actually be observed experimentally or not. It may be useful to emphasize this point, given that much of the previous literature on weak values focuses on their experimental observation in weak measurements and on the anomalous results that can be observed when unlikely measurement outcomes are post-selected \cite{Aha88}. In particular, I would like to avoid the misconception that weak values are defined operationally as the outcomes of weak measurements between state preparation and post-selection. This idea is highly problematic because it is not possible to satisfy the condition of infinitely weak measurement interactions in any real experiment. A fundamental and unambiguous definition of weak values needs to refer to the operator algebra of quantum mechanics, where the weak values represent the most natural assignment of a value to the operator $\hat{A}$ under the conditions $\mid \psi \rangle$ and $\mid m \rangle$. Their possible observation in weak measurements is merely an experimental confirmation of their essential role in the physical explanation of the randomness in the statistics of the final measurement outcomes $\mid m \rangle$. It is of course possible to argue that the outcomes of weak measurements alone establish the physical relevance of weak values, and it is significant that the anomaly of weak values itself can be used to prove contextuality \cite{Pus14}. However, the present discussion focuses on the theoretical definition of the weak value given by Eq.(\ref{eq:WV}). This definition provides a reliable estimate of the value of $\hat{A}$ under the conditions determined by state preparation $\mid \psi \rangle$ and measurement outcome $\mid m \rangle$, independent of any additional experimental evidence.

The drawback of using weak measurements to establish weak values is that the meter fluctuations in a weak measurement necessarily exceed the differences between the outcomes $A(m)$ associated with the post-selected final measurements $\mid m \rangle$. It is therefore difficult to tell whether the weak values are precise contextual values of $\hat{A}$ or not. The application of weak values as estimates of $\hat{A}$ can answer this question. As indicated by Eq.(\ref{eq:cqf}), weak values $A(m)$ provide a complete explanation of the uncertainty $\Delta A$ in the initial state $\mid \psi \rangle$ for any possible measurement $\{ \mid m \rangle \}$. The assignment of weak values therefore leaves no room for any additional uncertainties, proving that weak values are an error free estimate of $\hat{A}$ for any set of initial and final conditions expressed by pure states. This observation is consistent with the definitions of measurement errors introduced by Ozawa \cite{Oza03,Hal04}. Specifically, the error of the complex estimate $A(m)$ can be defined by the non-hermitian operator
\begin{equation}
\hat{\eta}_A = \hat{A} - \sum_m A(m) \mid m \rangle\langle m \mid.
\end{equation} 
If the complex weak values given by Eq. (\ref{eq:WV}) are used as the estimates $A(m)$, the initial state $\mid \psi \rangle$ is an eigenstate of the error operator $\hat{\eta}_A$ with an eigenvalue of zero,
\begin{equation}
\hat{\eta}_A \mid \psi \rangle = 0. 
\end{equation}
This eigenvalue equation indicates that the errors of the estimates $A(m)$ are precisely zero in the pure state $\mid \psi \rangle$. Since the operator $\hat{\eta}_A$ describes the difference between the operator $\hat{A}$ and the estimate $A(m)$, the quantum fluctuations of $\hat{A}$ in $\mid \psi \rangle$ should be equivalent to the quantum fluctuations of $A(m)$ associated with the statistics of measurement outcomes $\mid m \rangle$ in the same state. 

The situation changes considerably when we consider the quantum fluctuations of a mixed state $\hat{\rho}$. In this case, the residual error of the optimal estimate $A(m)$ does not vanish and the uncertainty $\Delta A$ is usually larger than the fluctuations of $A(m)$. The residual error can be expressed in terms of operator algebra \cite{Oza03,Hal04},
\begin{equation}
\label{eq:eta}
\eta_A^2 = \sum_m \langle m \mid (\hat{A}-A^*(m)) \,\hat{\rho}\, (\hat{A}-A(m)) \mid m \rangle.
\end{equation}
The optimal estimate is given by the weak values of the mixed state,
\begin{equation}
\label{eq:mixAm}
A(m) = \frac{\langle m \mid \hat{A}\, \hat{\rho}\mid m \rangle}{\langle m \mid \hat{\rho}\mid m \rangle},
\end{equation}
where the mixed state represents additional fluctuations of $\hat{A}$ around the statistical average of $A(m)$. These fluctuations can be quantified by using the contribution of the measurement outcome $m$ to the total error in Eq.(\ref{eq:eta}),
\begin{equation}
\label{eq:residual}
\eta_A^2(m) = \frac{\langle m \mid \hat{A} \,\hat{\rho} \hat{A}\, \mid m \rangle}{\langle m \mid \hat{\rho}\mid m \rangle} - \left|\frac{\langle m \mid \hat{A}\, \hat{\rho}\mid m \rangle}{\langle m \mid \hat{\rho}\mid m \rangle} \right|^2.
\end{equation} 
This error suggests that $A(m)$ represents an average of fluctuating values that are more precise than $A(m)$ because they include information about the ``classical'' fluctuations expressed by the mixed state. However, it is impossible to assign precise values to these fluctuations when no additional information about the initial state is available. However, such additional information might be available somewhere, and it might result in very different decompositions of the mixed state $\hat{\rho}$ into pure states $\mid \psi_\nu \rangle$. The residual quantum fluctuations of the mixed state $\hat{\rho}$ given by Eq.(\ref{eq:residual}) must be consistent with any possible decomposition into pure states $\mid \psi_\nu \rangle$ since it cannot be ruled out that such information is available somewhere. If the mixed state is expressed by a specific mixture of pure states, its statistics can be expressed by
\begin{equation}
\hat{\rho} = \sum_\nu \rho_\nu \mid \psi_\nu \rangle \langle \psi_\nu \mid.
\end{equation}
The weak value $A(m)$ is then given by the average of the weak values of $\mid \psi_\nu \rangle$ associated with this pure state decomposition. The conditional probabilities of each contribution $\nu$ are found by using the joint probabilities of $\nu$ and $m$ obtained from the sequential product of $\rho_\nu$ and $|\langle m \mid \psi_\nu \rangle|^2$ that appears in the total probability of $m$,
\begin{equation}
P(\nu|\hat{\rho},m) = \frac{\rho_\nu|\langle m \mid \psi_\nu \rangle|^2}{\langle m \mid \hat{\rho} \mid m \rangle}.
\end{equation}
The best estimate $A(m)$ is given by
\begin{equation}
A(m) = \sum_\nu \frac{\rho_\nu|\langle m \mid \psi_\nu \rangle|^2}{\langle m \mid \hat{\rho} \mid m \rangle}\; \frac{\langle m \mid \hat{A} \mid \psi_\nu \rangle}{\langle m \mid \psi_\nu \rangle},
\end{equation}
which is the average of the complex weak values associated with the different initial states $\mid \psi_\nu \rangle$ for the probability distribution $P(\nu|\hat{\rho},m)$ over $\nu$ conditioned by the initial mixed state and the final measurement outcome. This average over different pure state weak values can also explain the error of $A(m)$ given in Eq.(\ref{eq:residual}),
\begin{equation}
\label{eq:etafluct}
\eta_A^2(m) = \left(\sum_\nu \frac{\rho_\nu|\langle m \mid \psi_\nu \rangle|^2}{\langle m \mid \hat{\rho} \mid m \rangle}\left|\frac{\langle m \mid \hat{A} \mid \psi_\nu \rangle}{\langle m \mid \psi_\nu \rangle}\right|^2 \right) - \left|A(m)\right|^2.
\end{equation}
The error given by Eq.(\ref{eq:residual}) can therefore be explained in terms of the classical statistics described by the conditional probabilities $P(\nu|\hat{\rho},m)$. Importantly, {\it any} decomposition of the mixed state $\hat{\rho}$ into a mixture of pure states $\mid \psi_\nu \rangle$ can explain the residual quantum fluctuations given by the error $\eta_A^2(m)$ equally well, even though the weak values will vary greatly depending on the specific choice of the states. It would therefore be wrong to claim that the classical nature of the statistics justifies any kind of hidden variable model. Although the weak values of specific states $\nu$ can explain the conditional quantum fluctuations described by the error $\eta_A(m)$, there are no observable consequences of the individual values of $\hat{A}$ that could be obtained from the system itself after the measurement of $\mid m \rangle$ has been performed.

The problem posed by the absence of additional information about the mixed state $\hat{\rho}$ can be solved experimentally by using entangled states. If the mixed state $\hat{\rho}$ is the local density operator of a pure state that is entangled with a quantum mechanical reference any measurement of the reference will result in conditional states of the system, where the mixture of the conditional states is always given by the density matrix obtained by tracing out the reference. However, the different decompositions of the density matrix will not be compatible with each other due to the non-local selection of particular quantum coherences by the measurement of the reference. This non-local feature of entanglement is also known as steering \cite{Wis07}. In the following, I will investigate the effects of steering on the explanation of quantum fluctuations by weak values. It is then possible to identify the contextuality of quantum fluctuations in a mixed state by explaining the same initial uncertainty of $\hat{A}$ by different sets of weak values, where the context of quantum state preparation depends on the measurements made on the remote reference system.

\section{Quantum steering of weak values}
\label{Sec:steering}

In principle, any mixed state $\hat{\rho}$ can be extended into a pure entangled state $\mid E \rangle_{SR}$ by using the eigenstates $\mid \lambda \rangle$ of the density operator $\hat{\rho}$ to define the Schmidt decomposition of the pure state,
\begin{equation}
\mid E \rangle_{SR} = \sum_\lambda \sqrt{\rho_\lambda} \mid \lambda \rangle_S \otimes \mid \lambda \rangle_R.
\end{equation}
If we can prepare such a pure state entanglement, the mixed state $\hat{\rho}$ of the system will be decomposed into conditional pure states upon any precise projective measurement of the reference $R$. For an orthogonal measurement basis $\{\mid \nu \rangle_R\}$, the conditional states are given by
\begin{equation}
\sqrt{\rho_\nu} \mid \psi_\nu \rangle_S = \sum_\lambda \sqrt{\rho_\lambda} \langle \nu \mid \lambda \rangle \mid \lambda \rangle.
\end{equation}
The steering effect is clearly visible in the generation of coherences between the different eigenstates $\mid \lambda \rangle$ of the density operator $\hat{\rho}$. It is possible to suppress these coherences completely by choosing to measure $\{\mid \lambda \rangle\}$ in the reference system. In that case, the eigenstates of the density matrix $\hat{\rho}$ can be distinguished based on the measurement outcome obtained from the reference. Oppositely, any measurement of the reference represented by superpositions of $\mid \lambda \rangle$ will result in conditional states represented by corresponding superpositions in the system. 

The weak values associated with a particular measurement $\{\mid \nu \rangle_R\}$ of the reference can be expressed in terms of the pure states $\mid \psi_\nu \rangle_S$ conditioned by the measurement outcomes $\mid \nu \rangle_R$ obtained from the reference. It is possible to express these weak values in terms of the weak values associated with the eigenstates $\mid \lambda \rangle$ of the density operator, 
\begin{equation}
\label{eq:steer}
\frac{\langle m \mid \hat{A} \mid \psi_\nu \rangle}{\langle m \mid \psi_\nu \rangle}
= \sum_{\lambda} \frac{\langle m \mid \lambda \rangle \langle \lambda \mid \psi_\nu \rangle}{\langle m \mid \psi_\nu \rangle} \frac{\langle m \mid \hat{A} \mid \lambda \rangle}{\langle m \mid \lambda \rangle}.
\end{equation}
Quantum steering can thus be described by a fully coherent transformation of the context $(\mid \lambda \rangle; \langle m \mid)$ represented by the eigenstates of the local density operator $\hat{\rho}$ into the context $(\mid \psi_\nu \rangle; \langle m \mid)$ associated with the quantum states remotely prepared by obtaining the measurement result $\mid \nu \rangle_R$ from the reference. It should be noted that the transformation given in Eq.(\ref{eq:steer}) is just the general relation between different weak values when either the initial or the final state defining the weak value is changed \cite{Hof12}. Importantly, the coefficients of the linear transformation that relates weak values from different contexts to each other are themselves given by weak values of projection operators $\mid \lambda \rangle \langle \lambda \mid$ for the initial state $\mid \psi_\nu \rangle$ and the post-selected outcome $\mid m \rangle$. The change in context is therefore represented by weak values of conditional probabilities relating the part of the context that will be changed to the original context. This weak value analogy between the coefficients of a change in context and conditional probabilities means that Eq.(\ref{eq:steer}) looks very similar to the description of a statistical scattering process, despite the fact that it represents a fully reversible unitary transformation relating the information that can be obtained in a measurement of $\{\mid \lambda \rangle_R\}$ to the information that can be obtained in a measurement of $\{\mid \nu \rangle_R\}$. As shown in \cite{Hof15}, the negative real parts in the weak values defining the relations between different contexts explain quantum paradoxes in terms of the negative valued joint statistics of measurement contexts that never occur jointly. In particular, the steering effect on weak values can then be used to explain the violation of Bell's inequalities, as has already been demonstrated experimentally by Higgins et al. in \cite{Hig15}. In the light of the present discussion, this experiment provides a practical example of the steering effect on weak values, demonstrating how the control of the state preparation context by measurements of the reference steers the values assigned to projection operators between positive and negative values, resulting in correlations that cannot be represented by positive joint probabilities.

We can now analyze the contextuality of the quantum fluctuations given by the density matrix $\hat{\rho}$ in a two step process, where the first step consists in a measurement dependent estimate $A(m)$ as given by Eq.(\ref{eq:mixAm}). This estimate has an error of $\eta_A(m)$ as given by Eq.(\ref{eq:residual}), indicating that there are still residual quantum fluctuations left after the measurement context has been determined. In the second step, these quantum fluctuations can be resolved by a measurement of the entangled reference $R$. This measurement defines the individual values of the fluctuations as shown in Eq.(\ref{eq:steer}). By combining the two steps, we can express the complete quantum fluctuations of the initial mixed states in terms of a probability distribution over contexts defined by measurement outcomes $m$ and remote state preparations outcomes $\nu$,
\begin{equation}
\Delta A^2 = \left(\sum_{m,\nu} |\langle m; \nu \mid E \rangle|^2 \left|\frac{\langle m \mid \hat{A} \mid \psi_\nu \rangle}{\langle m \mid \psi_\nu \rangle} \right|^2\right) - (\mbox{Tr}(\hat{\rho} \hat{A}))^2.
\end{equation}
The application of remote state preparation highlights the symmetry between the measurement dependent part of the context and the state preparation part of the context. In the case of a mixed state, the state preparation part of the context is not completely determined. Likewise, additional measurement errors create an ambiguity in the measurement context. Statistical results with large amounts of excess noise are consistent with a wide variety of possible microscopic contexts so that all experimentally observable effects can be explained equally well by any of these contexts. However, it should be noted that additional information might invalidate most of these contexts. It is therefore important to remember that explanations are not necessarily correct just because they appear to fit the observable facts. In particular, we should not attribute reality to a specific decomposition of a mixed state just because of its convenient mathematical properties.

\section{Relations between different contexts}
\label{Sec:relations}

The analysis above is quite general and shows that the relation between different contexts is a fundamental feature of the Hilbert space formalism and its operator algebra. The most significant insight that should be gained from the use of steering to modify the quantum coherence of the input state is that the initial state preparation is also part of the context. This result can be illustrated by considering the case where the initial states are eigenstates of $\hat{A}$. In close analogy with the classical EPR argument \cite{EPR}, it is tempting to assume that the preparation of an eigenstate $\mid a \rangle$ of $\hat{A}$ means that the corresponding eigenvalue $A_a$ is now an element of reality, and not merely a part of the context established in quantum state preparation. This assumption can be justified by the weak values of $\hat{A}$, which are now equal to the eigenvalue $A_a$ for any measurement outcome $m$ of any measurement $\{ \mid m \rangle \}$. The simplicity of this argument makes it very tempting to prefer eigenstates of $\hat{A}$ to any other decomposition of a mixed state. If the density operator commutes with $\hat{A}$, it is possible to express $\hat{\rho}$ as a mixture of eigenstates $\mid a \rangle$,
\begin{equation}
\hat{\rho}=\sum_a \rho_a \mid a \rangle \langle a \mid.
\end{equation}
The fluctuations of $\hat{A}$ are given by the probability distribution $P(a)=\rho_a$, with
\begin{equation}
\Delta A^2 = \left(\sum_a \rho_a A_a^2\right) - \left(\sum_a \rho_a A_a\right)^2. 
\end{equation}
The best estimate of $\hat{A}$ for a measurement outcome of $m$ is given by the average over the conditional probabilities $P(a|\hat{\rho},m)$ for the different eigenvalues $A_a$ of the operator $\hat{A}$,
\begin{equation}
A(m) = \sum_a \frac{|\langle m \mid a \rangle|^2 \rho_a}{\langle m \mid \hat{\rho} \mid m \rangle} A_a.
\end{equation}
Here, the conditional probabilities of $a$ correspond to the conventional Bayesian results obtained from a prior of $P(a)=\rho_a$ and measurement probabilities of $P(m|a)=|\langle m \mid a \rangle|^2$. The residual quantum fluctuations of $\hat{A}$ for a measurement outcome of $m$ can also be expressed in terms of these probabilities,
\begin{equation}
\label{eq:Anoise}
\eta_A^2(m) = \left(\sum_a \frac{|\langle m \mid a \rangle|^2 \rho_a}{\langle m \mid \hat{\rho} \mid m \rangle} A_a^2\right) - (A(m))^2.
\end{equation}
As expected, quantum statistics reproduces the results of classical statistics whenever the quantum state $\hat{\rho}$ commutes with the target observable $\hat{A}$. However, it would be a fallacy to interpret this formal statistical correspondence as evidence that the eigenstates of $\hat{A}$ are automatically part of the context determined by the mixed state $\hat{\rho}$. As shown above, different decompositions of the mixed state are possible, and entanglement makes all of them experimentally accessible. 

Let us consider a reference measurement of $\{\mid \nu \rangle_R\}$ that is different from the eigenstates $\{\mid \lambda \rangle\} = \{\mid a \rangle\}$ of the density operator $\hat{\rho}$. The state preparation context is then decomposed into coherent superpositions of the eigenstates of $\hat{A}$ given by
\begin{equation}
\sqrt{\rho_\nu} \mid \psi_\nu \rangle_S = \sum_a \sqrt{\rho_a} \langle \nu \mid a \rangle \mid a \rangle.
\end{equation}
Eq. (\ref{eq:steer}) expresses the weak values associated with these remotely prepared states in terms of the eigenvalues of $\hat{A}$,
\begin{equation}
\frac{\langle m \mid \hat{A} \mid \psi_\nu \rangle}{\langle m \mid \psi_\nu \rangle}
= \sum_{a} \frac{\langle m \mid a \rangle \langle a \mid \psi_\nu \rangle}{\langle m \mid \psi_\nu \rangle} A_a.
\end{equation}
This is simply the conventional definition of weak values for initial states $\mid \psi_\nu \rangle$, expressed in terms of the complex conditional probabilities that describe the unitary transformations between different contexts \cite{Hof11,Hof12}. Significantly, the transformation from eigenvalues to weak values does not change the magnitude of the residual quantum fluctuations. The fluctuations of weak values obtained for the remotely prepared states $\mid \psi_\nu \rangle$ are expressed by the same error measure $\eta_A^2(m)$ defined by the eigenvalues of $\hat{A}$ in Eq.(\ref{eq:Anoise}) above. In the new state preparation context, this fluctuation of $\hat{A}$ can be expressed as
\begin{equation}
\label{eq:Bnoise}
\eta_A^2(m) = \left(\sum_\nu \frac{|\langle m \mid \psi_\nu \rangle|^2 \rho_\nu}{\langle m \mid \hat{\rho} \mid m \rangle} \left|\frac{\langle m \mid \hat{A} \mid \psi_\nu \rangle}{\langle m \mid \psi_\nu \rangle}\right|^2\right) - (A(m))^2.
\end{equation}
The equivalence of Eqs.(\ref{eq:Anoise}) and (\ref{eq:Bnoise}) has important implications for the physics of statistical fluctuations in mixed states. The possibility of obtaining different kinds of information about the fluctuations by choosing different kinds of measurements performed on the reference means that there is no valid interpretation of the mixed state as a mixture of pure states. The additional noise of a mixed state can only be described by classical statistics if the context of the missing information has been determined in a measurement of the entangled environment. In particular, this indicates that the solution of the measurement problem by decoherence is based on a fallacy. The mere possibility of decomposing a mixed state into a particular set of pure states is not sufficient to attribute any physical reality to these states. Contextuality must be taken seriously, since it provides the only microscopic description that is consistent with all experimental possibilities. 

The discussion above shows that eigenvalues are also contextual, even though they are the only contextual values that can be defined by state preparation or measurement alone. A practical example is the dominance of energy eigenstates in thermal statistics. It is tempting to interpret a thermal state as a mixture of energy eigenstates, but it is possible to obtain additional time dependent information from correlations between the system in a thermal state and other systems that interacted with it during the thermalization process. It is therefore necessary to accept that the energy uncertainty of thermal states can be quantum mechanical in nature, describing the time dependence of thermal diffusion processes in the quantum limit. In general, it is important to recognize that all decoherences leave microscopic traces of the original quantum coherence in the environment. The practical difficulty of recovering these traces should not be used as an argument to justify misinterpretations of the available data that cannot be upheld in situations where the level of control is improved. 

State preparation can determine one set of physical properties, and measurements can determine another. All remaining physical properties can be expressed as functions of the physical properties determined in this direct manner \cite{Hof12,Dir45}. The deviations between weak values and eigenvalues are a necessary consequence of the relation between physical properties described by the operator formalism \cite{Hof14,Nii18,Hof20b}. The correct explanation for this deviation is the contextuality of quantum fluctuations. Specifically, coherent superpositions express a fundamental kind of randomness that originates from the dynamics of state preparation and measurement. Quantum mechanical interactions leave no trace of the fluctuating properties in the environment, and that is the reason why it is possible to assign different sets of values to these fluctuations. The final measurement defines the values by eliminating all other effects that these quantum fluctuations might be having elsewhere through the disturbance of the state caused by the measurement. The reason why weak measurements can recover contextual values is because their measurement interaction is too weak to establish its own context. This means that weak measurements do not provide any evidence for the individual values of quantum fluctuations. Instead, the value is obtained by averaging over a large ensemble of systems prepared and measured in the same context. 

The discussion above shows what happens when additional information about the state preparation process becomes available through quantum correlations with a reference. This additional information updates the state preparation process and identifies the weak values obtained for a specific measurement outcome $m$ as averages of fluctuating values depending on the precise measurement performed to extract the additional information from the reference. Importantly, it is impossible to distinguish different decompositions of the quantum fluctuations on the system side. Any weak measurement evidence is based on the statistical average of large ensembles selected according to the measurement outcomes of the reference. However, the different weak values obtained by subdividing the events into ensembles associated with different reference outcomes $\nu$ provide a valid resolution of the residual uncertainty of the physical property $\hat{A}$ caused by the excess noise of the mixed state $\hat{\rho}$. If $\hat{\rho}$ and $\hat{A}$ commute, it may be tempting to apply a ``classical'' interpretation of noise whereby it is assumed that the statistical weight $\rho_A$ of the eigenstate $\mid a \rangle$ represents the initial probability of the eigenvalue $A_a$. However, the analysis above shows that this is an over interpretation of the diagonal form of the density matrix. The operator algebra of quantum fluctuations does not distinguish between the selection of weak values and the selection of eigenvalues. Both represent equally valid results associated with different measurements of the reference, and both cases fully explain the fluctuations $\eta_A(m)$ associated with the mixed state $\hat{\rho}$ and the measurement outcome $m$, as shown by the equivalence of the averaged squares, 
\begin{equation}
\sum_\nu \frac{|\langle m \mid \psi_\nu \rangle|^2 \rho_\nu}{\langle m \mid \hat{\rho} \mid m \rangle} \left|\frac{\langle m \mid \hat{A} \mid \psi_\nu \rangle}{\langle m \mid \psi_\nu \rangle}\right|^2 = \sum_a \frac{|\langle m \mid a \rangle|^2 \rho_a}{\langle m \mid \hat{\rho} \mid m \rangle} A_a^2.
\end{equation}
It might be useful to make the connection with energy fluctuations of a thermal state more explicit. The energy fluctuations play an important role in the exchange of energy between reservoirs, but their individual values have no impact on the observable thermodynamic effects. Quantum theory suggests that the dynamics of thermal fluctuations are expressed by superpositions of energy eigenstates, so the observation of time dependent effects in thermal states makes it impossible to identify energy eigenstates. It is nevertheless possible to assign weak values of energy, and these weak values result in the same magnitude of the average energy and its fluctuation as the assignment of eigenstates and eigenvalues did. 

It cannot be stressed enough that mixed states appear to be consistent with a wide range of contexts. This does not mean that contextuality disappears when there is sufficient noise. It really means that we do not know enough about the context to identify it properly. If this ambiguity of the context is not kept in mind, it is easy to misinterpret the mathematical description provided by quantum mechanics by attributing too much reality to elements of the formalism that only make sense in a specific context. A particular striking example for this fallacy is the tendency to attribute ``reality'' to eigenstates and eigenvalues, especially when it comes to properties such as energy or particle number. The discussion above clarifies how state preparation and measurement combine to objectively define the context based on the actual physical interactions with the system. It this provides a useful template for more thorough scientific discussions of all experimentally observable quantum phenomena.

\section{Conclusions}
\label{Sec:concl}

The discussion above has shown that the quantum fluctuations of a physical property $\hat{A}$ in an initial state can be resolved by optimal estimates $A(m)$ for each measurement outcome $m$. The optimal estimate is given by the complex weak values of the initial state and the measurement outcomes $m$. An estimation error of zero is obtained whenever the initial and the final state are pure states, indicating that any such combination establishes a precisely defined context for the quantum fluctuations of $\hat{A}$. On the other hand, mixed states do not define a precise context, leaving a part of the quantum fluctuations unresolved after the final measurement. The nature of these residual fluctuations can be analyzed using entanglement, where the remaining fluctuations can be resolved by measurements performed on the remote reference system. 

The analysis of quantum fluctuations conditioned by measurements of an entangled reference highlight the quantum nature of mixed state fluctuations, showing that even an apparent mixture of eigenstates of $\hat{A}$ should not be interpreted in terms of classical probabilities of the eigenvalues $A_a$ unless additional information about the precise eigenvalue $A_a$ can be obtained. Different contexts are only consistent when no further information is available. Contradictions between contexts can only be avoided if the physical incompatibility of different contexts is taken into account. 

Contextuality is not just an abstract notion used to explain away apparent contradictions between the statistics obtained in different measurements. The analysis presented above is intended to de-mystify contextuality and to pave the way towards a proper understanding of the manner in which quantum dynamics establishes the different contexts. In the present formulation, state vectors are used to summarize the complete physics of control involved in state preparation and quantum measurements. The results show that these processes play an important role in determining the objective properties of the system. Since state preparations must be consistent with all possible measurements, a quantum state can only define half of the context. Likewise, measurements must be consistent with all possible input states, so they too can only provide half of the context. Any violation of these limits would result in possible contradictions between the different contexts defined by preparation and measurement. Quantum fluctuations are needed to keep the context flexible enough. The use of entanglement highlights the symmetrical roles of state preparation and measurement, showing that the same measurement results can be explained by very different state preparation contexts. It is therefore important to consider the non-classical aspects of state preparation with as much care and attention as the problems associated with the fundamental limitations of quantum measurements. 


\end{document}